\documentclass[11pt]{article}

\usepackage{hyperref}  
\usepackage{graphicx}  
\usepackage{amssymb}   
\usepackage{amsmath}   

\title{\bf Critical coupling in \ensuremath{\phi_2^4} theory}

\author{Stephan D\"urr${}^{a,b}$ and Tolga S.\ H.\ Kiel${}^a$}

\date{${}^a${\sl Department of Physics, University of Wuppertal, 42119 Wuppertal, Germany}\\
${}^b${\sl J\"ulich Supercomputing Centre, Forschungszentrum J\"ulich, 52425 J\"ulich, Germany}}

\begin{document}

\maketitle

\begin{abstract}
We consider $\phi^4$ theory with $\phi(x)\in\mathbb{R}$ in two Euclidean dimensions. We determine for a variety of self-couplings $\hat{\lambda}$ the (negative) critical bare mass $\hat{\mu}_{0\mathrm{c}}^2(\hat{\lambda})$ where the lattice-regularized system changes from the symmetric to the broken phase. Based on these data, the transition to infinite volume and a universal scheme with the renormalized parameter $\hat{\mu}_\mathrm{c}^2(\hat{\lambda})$ is made. Finally, $f_\mathrm{c}=\lim_{\hat{\lambda}\to0}\hat{\lambda}/\hat{\mu}_\mathrm{c}^2(\hat{\lambda})$ is determined, with a judicious choice of the parameterizations considered. Our final result reads $f_\mathrm{c}=11.1097(20)_\mathrm{stat}(09)_\mathrm{sys}=11.1097(22)_\mathrm{tot}$.
\end{abstract}


\section{Introduction\label{sec:intro}}

The celebrated $\phi^4$-theory in two spacetime dimensions (2D) is the simplest conceivable quantum field theory.
In four spacetime dimensions (4D) the theory can be seen as the nucleus of the Higgs sector of the standard model in elementary particle physics.
In both 2D and 4D, the model shows a transition between a symmetric phase (where $\langle\sum\phi(x)\rangle=0$) and a broken phase (where $\langle\sum\phi(x)\rangle\neq0$) \cite{Simon:1973yz,Coleman:1985rnk}.
This transition has a non-perturbative origin, and it is technically challenging to locate the transition in the $(\lambda,\mu_0^2)$ parameter space [see (\ref{def_cont}) below],
and to measure physical properties such as the critical coupling $f_\mathrm{c}$ [as defined in (\ref{def_fc}) below].
In this paper we aim for an accurate determination of this quantity.

The theory in Euclidean space is defined by considering all possible field configurations $\phi(x)$ with $x\in\mathbb{R}^d$ and weighting them with the Boltzmann factor $\exp(-S)$ where
\begin{equation}
S=S[\phi(.)]=\int \frac{1}{2}\partial_\mu\phi(x)\partial_\mu\phi(x)+\frac{\mu_0^2}{2}\phi(x)^2+\frac{\lambda}{4}\phi(x)^4\;\mathrm{d}^dx
\label{def_orig}
\end{equation}
is the action of the field configuration.
With $[\phi]$ denoting the mass dimension of $\phi$, Eq.~(\ref{def_orig}) says $d=2[\phi]+2$ in $d$ spacetime dimensions.
Specifically, $[\phi]=0$ for $d=2$, and the bare parameters have mass dimensions $[\mu_0]=1$ and $[\lambda]=2$ in 2D.
On the lattice the integration $\int\mathrm{d}^dx$ is replaced by the discrete sum $a^d\sum_x$, where $a$ denotes the lattice spacing.
Consequently, in 2D the action reads
\begin{equation}
S=a^2\sum_{x\in\Lambda}\phi(x)\big(-\frac{1}{2}\triangle+\frac{\mu_0^2}{2}\big)\phi(x)+a^2\sum_{x\in\Lambda}\frac{\lambda}{4}\phi(x)^4
\label{def_cont}
\end{equation}
and with the discrete derivative $f''(x)=[f(x+a)-2f(x)+f(x-a)]/(2a^2)$ the action takes the form
\begin{equation}
S=\sum_{x\in\Lambda} \Big\{ -\sum_{\mu=1:2} \phi(x)\phi(x+\hat{\mu}) + \big(2+\frac{\hat{\mu}_0^2}{2}\big)\phi(x)^2 + \frac{\hat{\lambda}}{4}\phi(x)^4 \Big\}
\label{def_final}
\end{equation}
where $x+\hat{\mu}=x+ae_\mu$ is the neighboring site in the $\mu$-direction.
In addition, in Eq.~(\ref{def_final}) we introduced dimensionless siblings of the bare mass and self-coupling
\begin{equation}
\hat{\mu}_0=a\mu_0 \qquad\mbox{and}\qquad \hat{\lambda}=a^2\lambda\;.
\end{equation}

Given that $\lambda$ has a positive mass dimension in 2D, the $\phi^4$-theory in 2D is super-renormalizable; which means that only a finite number of diagrams must be considered in the renormalization process \cite{Coleman:1985rnk}.
For the model at hand, the details were worked out long ago in Refs.~\cite{Loinaz:1997az,Schaich:2009jk}.
The bottom line is that the renormalized $\hat\mu^2$, which corresponds to the bare $\hat\mu_0^2$, is given by the implicit solution of the system
\begin{align}
\hat{\mu}^2&=\hat{\mu}_0^2+3\hat{\lambda}A(\hat{\mu}^2)
\label{implicit_1}
\\
A(\hat{\mu}^2)&=\int_0^\infty e^{-\hat{\mu}^2t} [e^{-2t}I_0(2t)]^2 \,\mathrm{d}t
\label{implicit_2}
\;,
\end{align}
and the (numerically determined) bare $\hat{\mu}_{0\mathrm{c}}^2(\hat{\lambda})$ can be transcribed into a physical $\hat{\mu}_\mathrm{c}^2(\hat{\lambda})$.
On this basis one extracts
\begin{equation}
f_\mathrm{c} \equiv
\lim_{\hat{\lambda}\to0}f_\mathrm{c}(\hat{\lambda})
\qquad\mbox{with}\qquad
f_\mathrm{c}(\hat{\lambda})\equiv\frac{\hat{\lambda}}{\hat{\mu}_\mathrm{c}^2(\hat{\lambda})}
\label{def_fc}
\end{equation}
which is independent of the regularization used in the computation of $\hat{\mu}_{0\mathrm{c}}^2$ \cite{Loinaz:1997az,Schaich:2009jk}.

In this article we report an accurate numerical determination of $f_\mathrm{c}$ by Monte Carlo simulations.
For a large number of square lattice sizes ($N_x=N_y$) and parameters $\hat{\mu}_0^2$, $\hat{\lambda}$ we determine the finite-volume version of the field susceptibility \cite{Loinaz:1997az,Schaich:2009jk}
\begin{equation}
\chi=\langle\Phi^2\rangle-\langle|\Phi|\rangle^2
\qquad\mbox{with}\qquad
\Phi=\sum_{x\in\Lambda}\phi(x)
\label{def_susc}
\end{equation}
and we extract $f_\mathrm{c}$ as defined in (\ref{def_fc}) by means of a three-step procedure.

The remainder of this paper is organized as follows.
In Sec.~\ref{sec:peak} we specify the algorithms used and discuss how we scan, for a given set of parameters $(N_x,\hat{\lambda}$), the bare $\hat{\mu}_0^2$ to locate the peak in the susceptibility.
In Sec.~\ref{sec:infvol} we extrapolate our $\hat{\mu}_{0\mathrm{c}}^2$ to infinite volume and convert them to $\hat{\mu}_\mathrm{c}^2$ and subsequently to $f_\mathrm{c}(\hat{\lambda})$.
The most innovative part of the article is the analysis in Sec.~\ref{sec:fc}, which yields $f_\mathrm{c}$ with good control over all statistical and systematic uncertainties involved.
Our final result is compared with previous results in Sec.~\ref{sec:conc}, while technical details of the renormalization process are reviewed in App.~\ref{sec:app}.


\section{Determination of the peak positions in finite volume\label{sec:peak}}

Generically in lattice field theory, there is an algorithmic issue in the vicinity of a phase transition.
Specifically for the $\phi^4$-theory it was noticed that an embedded cluster-update step \cite{Brower:1989mt} (of Swendsen-Wang or Wolff type) is crucial for making significant moves in the phase space \cite{Loinaz:1997az,Schaich:2009jk}.
We combine several algorithms and define an ``update package'' that consists of a multi-hit Metropolis sweep, six overrelaxation sweeps, an HMC sweep, a random sign change of $\phi(x)$ across the lattice, a parity operation  on $\phi(x)$, and two embedded Wolff steps.
We chose to separate two consecutive measurements by $n_\mathrm{sepa}=10$ such ``update packages''.
Consequently, our measurements are reasonably well de-correlated, but we process our data by means of a jackknife analysis with sufficiently large blocksize, and we determine (as a check) the normalized autocorrelation functions of our observables.

\begin{figure}[!tb]
\includegraphics[width=0.5\linewidth]{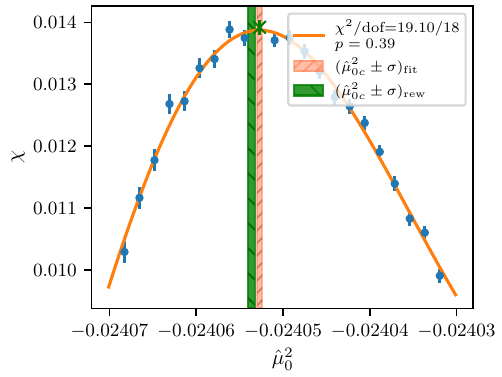}%
\includegraphics[width=0.5\linewidth]{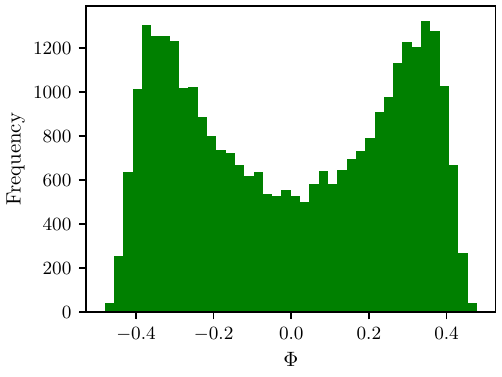}%
\caption{\label{fig:peak}
Left: Field susceptibility $\chi$ versus $\hat{\mu}_0^2$ at $\hat{\lambda}=0.01$ from $21$ independent simulations on $7168\times7168$ lattices. The peak position (with uncertainty) is shown by the orange vertical band. The closest simulation (marked by a cross) is reweighted to give a second determination (with uncertainty as indicated by the green vertical band). Right: Histogram of the field variable $\Phi$, as defined in (\ref{def_susc}), in the simulation that was marked with a green cross in the left panel.}
\end{figure}

It is impossible to determine by pure thought which lattices $N_x \times N_x$ should be simulated for a given $\hat{\lambda}$ to avoid non-linear finite-volume effects.
One might think of asking $N_x^2\hat{\lambda}\gg1$, as this is the dimensionless version of $V\lambda\gg1$, where $V=N_x^2a^2$ is the volume (in physical units) of the square box.
As we shall see, our lattice sizes are large enough to satisfy this criterion.
We will choose $N_x$ sufficiently large to facilitate a linear extrapolation of $\hat{\mu}_{0\mathrm{c}}^2(N_x)$ under $N_x\to\infty$ (see Sec.~\ref{sec:infvol} for details), and this leads to $N_x=N_x(\hat{\lambda})$ that are larger than the argument above would suggest (see below).

In infinite volume there is a bare $\hat{\mu}_0^2$ where the system changes from the symmetric to the broken phase (there is no conflict with the Mermin-Hohenberg-Wagner-Coleman theorem, since the underlying symmetry is $\mathbb{Z}_2$-valued).
In a finite volume and for a given observable, there is a peak at a specific value $\hat{\mu}_{0\mathrm{c}}^2$ (which depends on both $\hat{\lambda}$ and $N_x$), and the peak becomes more pronounced as the volume increases.
We choose the field susceptibility $\chi$ as defined in (\ref{def_susc}) as our diagnostic tool, due to its superior sensitivity.
The variance of the action (known as the ``specific heat'') is a legitimate choice, too, but it comes with much larger error-bars.

In the left panel of Fig.~\ref{fig:peak} an example of a scan of $\chi$ versus $\hat{\mu}_0^2$ is shown, here at $\hat{\lambda}=0.01$ on $7168^2$ lattices.
The simulations are independent; the error-bars are thus uncorrelated.
The overall structure looks roughly Gaussian, but with a slight ``tilt'' (purely Gaussian fits gave inacceptable $\chi^2/\mathrm{dof}$).
After some exercise we found that ``Gaussian+linear'' $a\cdot N(x;\mu,\sigma)+b\cdot(x-\mu)$ is a good fit model (with $3+1=4$ parameters); in this case we find $\chi^2/\mathrm{dof}=19.10/18=1.06$ and thus $p=0.39$).
This peak position and its uncertainty are marked by the orange vertical band in the figure.
Given the narrow spacing in $\hat{\mu}_0^2$ of our simulations, there is a second option to determine the peak position.
We select the simulation closest to the peak position as identified by the fit (in the figure marked by a green cross).
This simulation is used as the base ensemble to determine $\hat{\mu}_{0\mathrm{c}}^2$ via Ferrenberg-Swendsen reweighting \cite{Ferrenberg:1988yz,Ferrenberg:1989ui}; the result is indicated by the green vertical band.
The two results are always consistent, but often the reweighting yields a smaller error-bar, and this is why we choose the latter determination.
We should add that the effective sample size after reweighting is always $>98\%$ (in most cases even $>99\%$), indicating that the reweighting-shift is so small that this is a safe procedure.
The right panel of Fig.~\ref{fig:peak} features the distribution of the field variable $\Phi$, as defined in (\ref{def_susc}), in the simulation that was marked with a cross in the left panel (before reweighting).
At the edge of the symmetric phase the distribution resembles a two peak structure where the two peaks are at the brink of merging into a single peak, and there is a ``seam'' that connects the two ``melting'' peaks.


\begin{table}[!tb]
\centering
\begin{tabular}{c|c@{\hskip 2pt}c@{\hskip 2pt}c@{\hskip 2pt}c@{\hskip 2pt}l}
    \hline
    $\hat{\lambda}$ & 512 & 768 & 1024 & 1280 & 1536 \\
    \hline
    1.00 & $-1.2698570(43)$  & $-1.2707280(47)$  & $-1.2711522(49)$  & $-1.2714254(51)$  & $-1.2715982(52)$  \\
    0.70 & $-0.9493887(37)$  & $-0.9501151(39)$  & $-0.9504786(40)$  & $-0.9507007(40)$  & $-0.9508447(42)$  \\
    0.50 & $-0.7193931(66)$  & $-0.7200037(63)$  & $-0.7203108(49)$  & $-0.7204902(47)$  & $-0.7206141(40)$  \\
    0.35 & $-0.5343740(44)$  & $-0.5348981(45)$  & $-0.5351463(50)$  & $-0.5352960(53)$  & $-0.5353999(61)$  \\
    0.25 & $-0.4025047(45)$  & $-0.4029331(40)$  & $-0.4031513(32)$  & $-0.4032841(28)$  & $-0.4033610(27)$  \\
    0.15 & $-0.2603995(29)$  & $-0.2607336(79)$  & $-0.2608988(72)$  & $-0.2610125(70)$  & $-0.2610826(73)$  \\
    0.10 & $-0.1835496(32)$  & $-0.1838206(34)$  & $-0.1839555(31)$  & $-0.1840360(26)$  & $-0.1840952(23)$  \\
    0.05 & $-0.1001814(18)$  & $-0.1003677(16)$  & $-0.1004614(15)$  & $-0.1005192(14)$  & $-0.1005561(13)$  \\
    0.03 & $-0.0637694(15)$  & $-0.0639119(14)$  & $-0.0639848(12)$  & $-0.0640299(12)$  & $-0.0640565(11)$  \\ 
    0.02 & $-0.04442676(41)$ & $-0.04454188(42)$ & $-0.04460053(43)$ & $-0.04463700(33)$ & $-0.04466107(32)$ \\
    0.01 & $-0.02382478(34)$ & $-0.02390340(34)$ & $-0.02394420(34)$ & $-0.02396929(26)$ & $-0.02398610(27)$ \\
     \hline
\end{tabular}
\\[4pt]
\begin{tabular}{@{\hskip 2pt}c|c@{\hskip 2pt}c@{\hskip 2pt}c@{\hskip 2pt}c@{\hskip 2pt}c@{\hskip 2pt}c@{\hskip 2pt}}
    \hline
    $\hat{\lambda}$ & 2048 & 3072 & 4096 & 5120 & 6144 & 7168 \\
    \hline
    1.00 & $-1.2718191(53)$  & x & x & x & x & x \\
    0.70 & $-0.9510199(45)$  & x & x & x & x & x \\
    0.50 & $-0.7207671(30)$  & x & x & x & x & x \\
    0.35 & $-0.5355304(82)$  & x & x & x & x & x \\
    0.25 & $-0.4034762(21)$  & x & x & x & x & x \\
    0.15 & $-0.2611629(46)$  & x & x & x & x & x \\
    0.10 & $-0.1841574(46)$  & $-0.1842285(15)$  & $-0.1842618(16)$  & $-0.1842824(16)$  & $-0.1842939(16)$  & x \\
    0.05 & $-0.1006039(10)$  & $-0.1006528(14)$  & $-0.1006772(13)$  & $-0.1006927(13)$  & x & x \\
    0.03 & $-0.06409552(87)$ & $-0.0641314(18)$  & $-0.06415165(83)$ & $-0.06416139(78)$ & x & x \\ 
    0.02 & $-0.04469033(99)$ & $-0.04472054(69)$ & $-0.0447369(12)$  & $-0.04474579(68)$ & x & x \\
    0.01 & $-0.02400694(25)$ & $-0.02402761(56)$ & $-0.02403958(53)$ & $-0.02404612(54)$ & $-0.02404969(48)$ & $-0.02405359(46)$ \\
     \hline
\end{tabular}
\caption{\label{tab:firststage}
Results for $\hat{\mu}_{0\mathrm{c}}^2(\hat{\lambda},N_x)$ from the first stage fits. For each $\hat{\lambda}$ and geometry $N_x\times N_x$ about $O(20)$ streams at individual $\hat{\mu}_0^2$ in the vicinity of the peak were simulated, see Fig.~\ref{fig:peak} for the actual simulations at $(0.01,7168)$.}
\end{table}

To give a more complete view of how many ensembles were generated for this article, Tab.~\ref{tab:firststage} lists the lattice sizes $N_x \times N_x$ that were simulated for each $\hat{\lambda}$.
The smaller $\hat{\lambda}$ the larger the box size $N_x$ needed to establish a linear extrapolation of $\hat{\mu}_{0\mathrm{c}}^2$ under $N_x\to\infty$ (see Sec.~\ref{sec:infvol} below).
For each set $(\hat{\lambda},N_x)$ the peak position is determined from approximately 20 ensembles, and each ensemble consists of a significant number of more-or-less decorrelated configurations [at least $10^4$ for $N_x\geq3072$, and at least $10^5$ for $N_x\leq2048$].
The peak position, as determined by the field susceptibility, is also listed in this table.


\section{Extrapolation to infinite volume and renormalization\label{sec:infvol}}

\begin{figure}[!tb]
\includegraphics[width=0.5\linewidth]{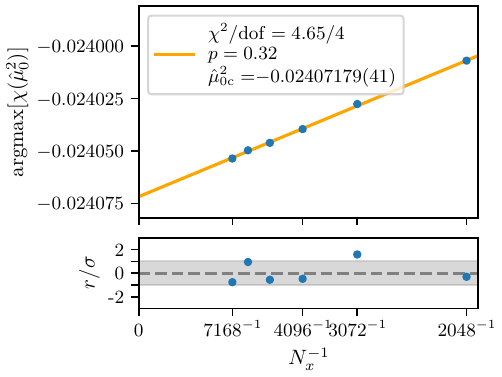}%
\includegraphics[width=0.5\linewidth]{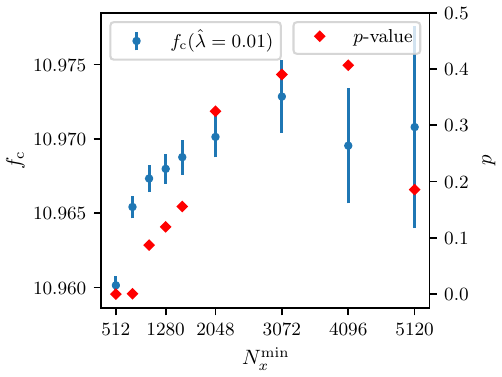}%
\caption{\label{fig:infvol}
Left: Extrapolation of the peak positions $\hat{\mu}_{0\mathrm{c}}^2(\hat{\lambda},N_x)$ at $\hat{\lambda}=0.01$ to $N_x\to\infty$ for $N_x\geq2048$, along with the resulting pulls (in standard deviations) of the respective data point. The data with $N_x=512,768,1024$ are not included in the fit and out of scale.
Right: $f_\mathrm{c}$ at $\hat{\lambda}=0.01$ (after loop-correction) versus the minimal $N_x$ included in the fit. The red diamonds indicate the resulting $p$-values.}
\end{figure}

With the critical $\hat{\mu}_0^2$-values for each $(\hat{\lambda},N_x)$ combination in hand, it is possible to eliminate the effect of the finite box-volume by taking the limit $N_x\to\infty$.
A typical extrapolation is shown in the left panel of Fig.~\ref{fig:infvol}, again in the case of our smallest $\hat{\lambda}$.
Due to our choice of simulating large $N_x\times N_x$ lattices, the extrapolation is linear in $1/N_x$.
The extrapolation shown involves the results for $\hat{\mu}_{0\mathrm{c}}^2$, as determined in Sec.~\ref{sec:peak}, with $N_x\geq2048$ (for $\hat{\lambda}=0.01$).
This produces a good fit quality (with $\chi^2/\mathrm{dof}=4.65/4$ or $p=0.32$), and the pulls show no obvious trend.
Of course, the extrapolated $\hat{\mu}_{0\mathrm{c}}^2$ changes slightly if one more $N_x$ is included in the fit (or excluded from the fit).
The variation of the final result (modulo a map $\hat{\mu}_{0\mathrm{c}}^2 \to f_\mathrm{c}$ to be discussed below) and of the $p$-value is shown in the right panel.
In the case shown, any of the choices $N_x^\mathrm{max}=2048,3072,4096$ yields a good $p$-value and an $f_\mathrm{c}$ that is consistent with all subsequent values.
In such a situation, one might opt for $N_x^\mathrm{max}=2048$ or average over the three possibilities; it hardly makes any difference.

\begin{table}[!tb]
\centering
\begin{tabular}{c|ccc}
\hline
$\hat{\lambda}$ & $\hat{\mu}_{0\mathrm{c}}^2(\hat{\lambda})$ & $\hat{\mu}_\mathrm{c}^2(\hat{\lambda})$ & $f_\mathrm{c}(\hat{\lambda})$\\
\hline
1.00 & $-1.2724682(48)$  & $0.0973330(13)$  & $10.27401(14)$ \\
0.70 & $-0.9515697(38)$  & $0.0684496(11)$  & $10.22650(16)$ \\
0.50 & $-0.7212248(39)$  & $0.0488540(11)$  & $10.23458(23)$ \\
0.35 & $-0.5359149(57)$  & $0.0340400(16)$  & $10.28203(49)$ \\
0.25 & $-0.4037982(27)$  & $0.02415421(76)$ & $10.35016(33)$ \\
0.15 & $-0.2614170(47)$  & $0.0143181(13)$  & $10.47623(97)$ \\
0.10 & $-0.18436329(97)$ & $0.00945429(27)$ & $10.57721(31)$ \\
0.05 & $-0.10074843(94)$ & $0.00465746(26)$ & $10.73547(61)$ \\
0.03 & $-0.06420590(66)$ & $0.00277030(18)$ & $10.82914(72)$ \\
0.02 & $-0.04478158(56)$ & $0.00183643(16)$ & $10.89067(92)$ \\
0.01 & $-0.02407179(41)$ & $0.00091157(11)$ & $10.9701(14)$  \\
\hline
\end{tabular}
\caption{\label{tab:infvol_limits}
Critical $\hat{\mu}_{0\mathrm{c}}^2(\hat{\lambda})$, $\hat{\mu}_\mathrm{c}^2(\hat{\lambda})$ and $f_\mathrm{c}(\hat{\lambda})\equiv\hat{\lambda}/\hat{\mu}_\mathrm{c}^2(\hat{\lambda})$, after extrapolation to $N_x\to\infty$, for different $\hat{\lambda}$.}
\end{table}

\begin{figure}[!tb]
\includegraphics[width=0.5\linewidth]{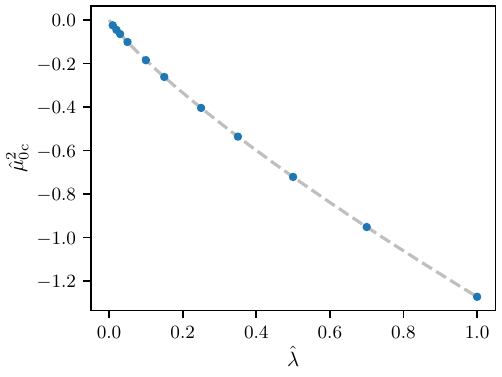}%
\includegraphics[width=0.5\linewidth]{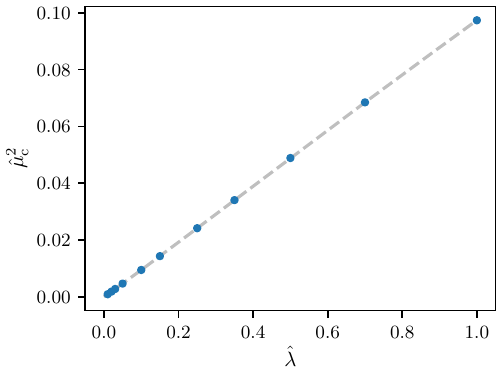}%
\caption{\label{fig:phases}
Phase diagrams of $\phi_2^4$ theory in terms of the bare $\hat{\mu}_{0\mathrm{c}}^2$ (left) and the renormalized $\hat{\mu}_{\mathrm{c}}^2$ (right).
In either panel cubic splines connect the points (whose error-bars are dwarfed by the symbol size), and the broken phase is beneath the line.
The desired quantity $f_\mathrm{c}$ is the inverse of the asymptotic slope (near the origin) of the transition line in the right panel.}
\end{figure}

The extrapolated $\hat{\mu}_{0\mathrm{c}}^2$ are listed in Tab.~\ref{tab:infvol_limits} as a function of $\hat{\lambda}$.
Connecting these points in the $(\hat{\lambda},\hat{\mu}_{0\mathrm{c}}^2)$ plane establishes a phase diagram of the theory in the bare parameter space, which shows where the system (in infinite volume) is in the symmetric phase and where the $\mathbb{Z}_2$ symmetry is broken.
This is shown in the left panel of Fig.~\ref{fig:phases}.

The next step is to trade $\hat{\mu}_{0\mathrm{c}}^2$ for $\hat{\mu}_\mathrm{c}^2$ by solving the implicit equation (\ref{implicit_1}).
Some background on how this equation was derived in Refs.~\cite{Loinaz:1997az,Schaich:2009jk} is arranged in App.~\ref{sec:app}.
The conversion is done with a few iteration steps and (\ref{implicit_2}) is inexpensive to evaluate.
This procedure converts the first to the second column in Tab.~\ref{tab:infvol_limits}.
Again, we might connect these points in the $(\hat{\lambda},\hat{\mu}_\mathrm{c}^2)$ plane to establish the phase diagram of the theory in the renormalized parameter space.
This is shown in the right panel of Fig.~\ref{fig:phases}.
Finally, by means of the definition $f_\mathrm{c}(\hat{\lambda})\equiv\hat{\lambda}/\hat{\mu}_\mathrm{c}^2(\hat{\lambda})$ the last column is obtained.

According to (\ref{def_fc}) the last missing step is the extrapolation of $f_\mathrm{c}(\hat{\lambda})$ to $\hat{\lambda}=0$.
Due to $\hat{\lambda}=a^2\lambda$ this is equivalent to the continuum limit.
This extrapolation is rather tricky, and this is why we shall discuss it in a separate section.


\section{Critical coupling in the continuum limit\label{sec:fc}}

From a look at Tab.~\ref{tab:infvol_limits} it is clear that extrapolating the rightmost column to $\hat{\lambda}=0$ is non-trivial.
The reason is the near-monotonic increase of the error-bar towards the desired limit (from 14 on the last digit to 140).
This means that the fit will be dominated by the precise data at large $\hat{\lambda}$, but we would like to learn something about the behavior at small $\hat{\lambda}$.

In addition, there is no god-given functional form that should be used to fit the data.
Historically, people started fitting the data with polynomials \cite{Loinaz:1997az}, but soon it was realized that a logarithmic contribution is needed to fit the (more precise) data \cite{Schaich:2009jk}.
In the absence of additional analytical guidance, we employ the ansatz
\begin{equation}
f_\mathrm{c}(\hat{\lambda})=\sum_{m,n\geq0}c_{mn}\hat{\lambda}^m\ln^n(\hat{\lambda})
\label{def_fits}
\end{equation}
with the condition that the combination $(m=0,n>0)$ shall be prohibited, to ensure a finite limit $\hat{\lambda}\to0$.
Beyond this we are unaware of a physics based argument why one should exclude further $(m,n)$ combinations.
In perturbation theory (in particular in perturbative QCD) it is natural to resum certain classes of diagrams, and one often ends up with N$^n$LL (``leading log'') results with high powers $\ln^n(\lambda)$ and modest powers $\lambda^m$.

\begin{figure}[!tb]
\includegraphics[width=0.5\linewidth]{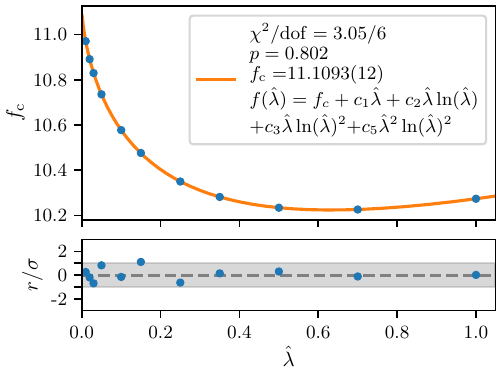}%
\includegraphics[width=0.5\linewidth]{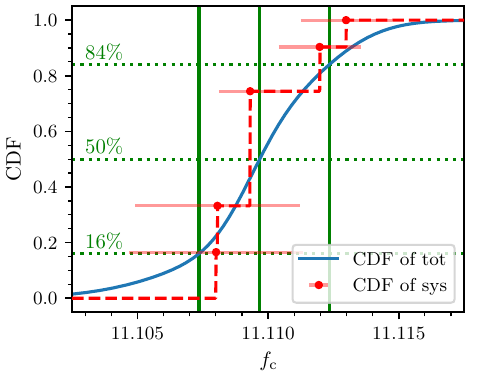}%
\caption{\label{fig:fc_lambdamax_1}
Left: Example of a continuum fit with $\hat{\lambda}_\mathrm{max}=1.0$. Right: Cumulative distribution function (CDF) of all statistically acceptable fits with $\hat{\lambda}_\mathrm{max}=1.0$, weighted with their AIC scores. At each step of the red dashed curve, the horizontal band indicates the statistical uncertainty of the fit. The blue curve shows a CDF obtained by combining the AIC weights with the Gaussian distributions corresponding to the statistical uncertainties.}
\end{figure}

We begin our discussion with fits that include all available $\hat{\lambda}$, so $\hat{\lambda}_\mathrm{max}=1$.
A representative of this class is shown in the left panel of Fig.~\ref{fig:fc_lambdamax_1}.
In fact, this is the best fit available for the full dataset, with a notable $\chi^2/\mathrm{dof}=3.05/6$ and thus $p=0.802$.
Moreover, the ``pulls'' (data-fit) show no obvious trend, and we conclude that this fit gives a very good description of the data.
The functional form is $f(\hat{\lambda})=f_\mathrm{c}+c_1\hat{\lambda}+c_2\hat{\lambda}\ln(\hat{\lambda})+c_3\hat{\lambda}\ln(\hat{\lambda})^2+c_5\hat{\lambda}^2\ln(\hat{\lambda})^2$,
and we have the covariance matrix of the five parameters involved, but physically relevant is only $f_\mathrm{c}=11.1093(12)$.
Note that this error-bar quotes only the statistical uncertainty (and it is specific to the fit discussed).

\begin{table}[!tb]
\centering
\begin{tabular}{@{\hskip 5pt}l@{\hskip 5pt}c@{\hskip 5pt}c@{\hskip 5pt}c@{\hskip 5pt}c@{\hskip 5pt}}
\hline
fit ansatz & $f_\mathrm{c}$ & $\chi^2/\mathrm{dof}$ & $p$ & $w_\mathrm{AIC}$ \\
\hline
$f(\hat{\lambda})=f_\mathrm{c}+c_1\hat{\lambda}+c_2\hat{\lambda}\ln(\hat{\lambda})$ & $10.89803(37)$ & 60000.0/8 & 0.00 & 0.00\\
\hline
$f(\hat{\lambda})=f_\mathrm{c}+c_1\hat{\lambda}+c_2\hat{\lambda}\ln(\hat{\lambda})+c_3\hat{\lambda}\ln(\hat{\lambda})^2$ & $11.10505(94)$ & 36.0/7 & 0.00 & 0.00 \\
$f(\hat{\lambda})=f_\mathrm{c}+c_1\hat{\lambda}+c_2\hat{\lambda}\ln(\hat{\lambda})+c_4\hat{\lambda}^2\ln(\hat{\lambda})$ & $10.99691(58)$ & 6380.0/7 & 0.00 & 0.00 \\
$f(\hat{\lambda})=f_\mathrm{c}+c_1\hat{\lambda}+c_2\hat{\lambda}\ln(\hat{\lambda})+c_5\hat{\lambda}^2\ln(\hat{\lambda})^2$ & $10.87917(40)$ & 42500.0/7 & 0.00 & 0.00 \\
$f(\hat{\lambda})=f_\mathrm{c}+c_1\hat{\lambda}+c_2\hat{\lambda}\ln(\hat{\lambda})+c_6\hat{\lambda}^2$ & $11.02459(66)$ & 3090.0/7 & 0.00 & 0.00 \\
\hline
$f(\hat{\lambda})=f_\mathrm{c}+c_1\hat{\lambda}+c_2\hat{\lambda}\ln(\hat{\lambda})+c_3\hat{\lambda}\ln(\hat{\lambda})^2+c_4\hat{\lambda}^2\ln(\hat{\lambda})$ & $11.1120(16)$ & 4.95/6 & 0.55 & 0.14 \\
$f(\hat{\lambda})=f_\mathrm{c}+c_1\hat{\lambda}+c_2\hat{\lambda}\ln(\hat{\lambda})+c_3\hat{\lambda}\ln(\hat{\lambda})^2+c_5\hat{\lambda}^2\ln(\hat{\lambda})^2$ & $11.1093(12)$ & 3.05/6 & 0.80 & 0.36 \\
$f(\hat{\lambda})=f_\mathrm{c}+c_1\hat{\lambda}+c_2\hat{\lambda}\ln(\hat{\lambda})+c_3\hat{\lambda}\ln(\hat{\lambda})^2+c_6\hat{\lambda}^2$ & $11.1130(17)$ & 5.96/6 & 0.43 & 0.08 \\
$f(\hat{\lambda})=f_\mathrm{c}+c_1\hat{\lambda}+c_2\hat{\lambda}\ln(\hat{\lambda})+c_4\hat{\lambda}^2\ln(\hat{\lambda})+c_5\hat{\lambda}^2\ln(\hat{\lambda})^2$ & $11.0564(10)$ & 306.0/6 & 0.00 & 0.00 \\
$f(\hat{\lambda})=f_\mathrm{c}+c_1\hat{\lambda}+c_2\hat{\lambda}\ln(\hat{\lambda})+c_4\hat{\lambda}^2\ln(\hat{\lambda})+c_6\hat{\lambda}^2$ & $11.0781(12)$ & 65.1/6 & 0.00 & 0.00 \\
$f(\hat{\lambda})=f_\mathrm{c}+c_1\hat{\lambda}+c_2\hat{\lambda}\ln(\hat{\lambda})+c_5\hat{\lambda}^2\ln(\hat{\lambda})^2+c_6\hat{\lambda}^2$ & $11.0646(10)$ & 191.0/6 & 0.00 & 0.00 \\
\hline
$f(\hat{\lambda})=f_\mathrm{c}+c_1\hat{\lambda}+c_2\hat{\lambda}\ln(\hat{\lambda})+c_3\hat{\lambda}\ln(\hat{\lambda})^2+c_4\hat{\lambda}^2\ln(\hat{\lambda})+c_5\hat{\lambda}^2\ln(\hat{\lambda})^2$ & $11.1081(31)$ & 2.86/5 & 0.72  & 0.14 \\
$f(\hat{\lambda})=f_\mathrm{c}+c_1\hat{\lambda}+c_2\hat{\lambda}\ln(\hat{\lambda})+c_3\hat{\lambda}\ln(\hat{\lambda})^2+c_4\hat{\lambda}^2\ln(\hat{\lambda})+c_6\hat{\lambda}^2$ & $11.1071(39)$ & 3.02/5 & 0.70  & 0.13 \\
$f(\hat{\lambda})=f_\mathrm{c}+c_1\hat{\lambda}+c_2\hat{\lambda}\ln(\hat{\lambda})+c_3\hat{\lambda}\ln(\hat{\lambda})^2+c_5\hat{\lambda}^2\ln(\hat{\lambda})^2+c_6\hat{\lambda}^2$ & $11.1080(33)$ & 2.87/5 & 0.72 & 0.14 \\
$f(\hat{\lambda})=f_\mathrm{c}+c_1\hat{\lambda}+c_2\hat{\lambda}\ln(\hat{\lambda})+c_4\hat{\lambda}^2\ln(\hat{\lambda})+c_5\hat{\lambda}^2\ln(\hat{\lambda})^2+c_6\hat{\lambda}^2$ & $11.0938(24)$ & 6.90/5 & 0.23  & 0.02 \\
\hline
$f(\hat{\lambda})=f_\mathrm{c}+c_1\hat{\lambda}+c_2\hat{\lambda}\ln(\hat{\lambda})+c_3\hat{\lambda}\ln(\hat{\lambda})^2+c_4\hat{\lambda}^2\ln(\hat{\lambda})+c_5\hat{\lambda}^2\ln(\hat{\lambda})^2+c_6\hat{\lambda}^2$ & $11.1129(93)$ & 2.57/4 & 0.63  & 0.06 \\
\hline
\end{tabular}
\caption{\label{tab:lambda_max_1.0}
Results of the fits based on the ``base model'' $f_\mathrm{c}+c_1\hat{\lambda}+c_2\hat{\lambda}\ln(\hat{\lambda})$ plus any of the terms with $c_3,c_4,c_5,c_6$ switched on or off. The grouping with horizontal lines is meant to indicate how many of these terms are switched on. The fitted $f_\mathrm{c}$ is given with its statistical uncertainty and the goodness of fit. All fits are for $\hat{\lambda}_\mathrm{max}=1.0$.}
\end{table}

This specific fit is the result of a systematic procedure that scans the available ans\"atze (\ref{def_fits}).
Starting from the fit ansatz $f(\hat{\lambda})=f_\mathrm{c}+c_1\hat{\lambda}+c_2\hat{\lambda}\ln(\hat{\lambda})$, as used in Ref.~\cite{Schaich:2009jk},
one might add a term $c_3\hat{\lambda}\ln(\hat{\lambda})^2$ or $c_4\hat{\lambda}^2\ln(\hat{\lambda})$ or $c_5\hat{\lambda}^2\ln(\hat{\lambda})^2$ or $c_6\hat{\lambda}^2$.
In fact, one might add any combination of these terms, hoping that this will improve the fit quality.

Although this is not the systematic procedure mentioned (which we shall specify soon), we feel it makes sense to consider all combinations of $2^4=16$ ans\"atze with $c_3,c_4,c_5,c_6$ switched on or off.
Specifically for $\hat{\lambda}_\mathrm{max}=1.0$ (that is the maximum range possible) this survey is presented in Tab.~\ref{tab:lambda_max_1.0}.
The first line contains the ``base model'' with none of these terms switched on; it yields $\chi^2\simeq60\,000$ with $\mathrm{dof}=11-3=8$ and thus an inacceptable $p$-value.
The next four lines contain the fits where \emph{one} of the terms discussed is switched on (i.e.\ the fits with $\mathrm{dof}=7$).
None of them has an acceptable $\chi^2/\mathrm{dof}$, but it is clear that the fit with $c_3$ works much better than the remaining three fits.
The following six lines contain the fits with \emph{two} of the terms switched on (i.e.\ the fits with $\mathrm{dof}=6$).
Three of them have good $\chi^2/\mathrm{dof}$, while the remaining three are inacceptable.
The key observation is that the three good fits all contain the $c_3$ term that did best at the previous level ($\mathrm{dof}=7$).
The next four lines contain the fits with \emph{three} of the terms switched on (i.e.\ the fits with $\mathrm{dof}=5$).
Here all four fits work well, but the three lines with $c_3$ are a bit more convincing than the line without the $c_3$-term.
The last line contains the fit with all \emph{four} terms switched on, and it would give an acceptable model.

The bottom line of this discussion is that there are eight fits in Tab.~\ref{tab:lambda_max_1.0} with acceptable $p$-value (say $p>0.05$ or similar) and eight more that can be ignored.
Hence, a natural way to proceed would be to form a weighted average of the central values, for example $f_\mathrm{c}=(0.55\cdot11.1120(16)+0.80\cdot11.1093(12)+0.43\cdot11.1130(17)+0.72\cdot11.1081(31)+0.70\cdot11.1071(39)+0.72\cdot11.1080(33)+0.23\cdot11.0938(24)+0.63\cdot11.1129(93))/(0.55+0.80+0.43+0.72+0.70+0.72+0.23+0.63)$.
This gives $f_\mathrm{c}=11.1090(34)_\mathrm{stat}$, where we emphasize that the uncertainty indicated is only statistical.

There are three measures to improve this procedure (before considering other choices than $\hat{\lambda}_\mathrm{max}=1.0$).
The first measure is ``pruning''.
The idea is to reduce the overhead spent on fits that do not work because their $p$-value is essentially zero.
In the example discussed, out of the four options for augmenting the ``base model'' by one term, the term with $c_3$ gave a much better $\chi^2/\mathrm{dof}$ than the other three terms ($36/7$ as opposed to $>3000/7$).
The idea of ``pruning'' is to select this $4$-parameter ansatz for further refinement, since the three fits in the middle six-fit block are all descendants of this ansatz (they all have the $c_3$-term).
In the next step the best $5$-parameter fit available (the one with $p=0.80$) is augmented by yet another term.
This yields the fits with $c_3$, $c_5$ and one more term, i.e.\ those with $p\in\{0.72,0.72\}$.
At this point the ``pruning'' stops, since each new $p$-value is below the value $p=0.80$ that was achieved in the previous step.
The pruning history thus leaves us with $p=0.0,0.0,0.55,0.80,0.43,0.72,0.72$, that is with five fits that contribute significantly.
Consequently, the weighted average simplifies to $f_\mathrm{c}=(0.55\cdot11.1120(16)+0.80\cdot11.1093(12)+0.43\cdot11.1130(17)+0.72\cdot11.1081(31)+0.72\cdot11.1080(33))/(0.55+0.80+0.43+0.72+0.72)$.

The second measure is to use the ``Akaike information criterion'' (AIC) \cite{Akaike:1974vps} for the weighting rather than the $p$-value.
The latter criterion is based on $\chi^2/(n-k)$, where $n$ is the number of data and $k$ is the number of parameters in the ansatz.
The former criterion uses the weights $w_\mathrm{AIC}=\exp{(-\frac{1}{2}[\chi^2+2k-n])}$, later normalized to fulfill $\sum w_\mathrm{AIC}=1$, and it is known to penalize any ``overfitting'' more vigorously.
The AIC is widely used in statistical analysis in many fields, including lattice field theory \cite{BMW:2014pzb,Borsanyi:2020mff}.

The third measure is to form the cumulative distribution function (CDF), and this is illustrated in the right panel of Fig.~\ref{fig:fc_lambdamax_1}.
The five fits emerging from the pruning procedure define five values of $f_\mathrm{c}$ where we install a step function whose height is proportional to the AIC-weight of the fit (the overall height sums up to $1$).
The resulting five-step curve is shown as a dashed red line, and the abscissa value $f_\mathrm{c}=11.1093$ where it exceeds $0.5$ is a reasonable estimate of the central value.
Alternatively, a normal cumulative distribution function (essentially an error function) could be fitted to the five-step function.
But it is legitimate to ``smooth out'' each step by replacing it with a normal cumulative distribution whose width is defined by the statistical uncertainty of the respective fit.
The resulting curve is shown with a full blue line and the positions where it reaches $0.16$ and $0.84$ yield $f_\mathrm{c}=11.1097(25)_\mathrm{tot}$.
For details of how we split this total error into its statistical and systematic contributions, we refer to Ref.~\cite{Borsanyi:2020mff} and App.~\ref{sec:app_b}; this gives $f_\mathrm{c}=11.1097(18)_\mathrm{stat}(18)_\mathrm{syst}$ (for $\hat{\lambda}_\mathrm{max}=1.0$).

\begin{table}[!tb]
\begin{minipage}{0.5\linewidth}%
\begin{tabular}{lcccc}
\hline
fit ansatz & $f_\mathrm{c}$ & $\chi^2/\mathrm{dof}$ & $p$ & $w_\mathrm{AIC}$ \\
\hline
base model & $10.95571(48)$ & 19600.0/7 & 0.00 & 0.00 \\
\hline
$\{c_3\}$ & $11.1094(12)$ & 2.86/6 & 0.83 & 0.65 \\
$\{c_4\}$ & $11.02283(70)$ & 2120.0/6 & 0.00 & 0.00 \\
$\{c_5\}$ & $11.0896(11)$ & 2400.0/6 & 0.00 & 0.00 \\
$\{c_6\}$ & $11.04947(84)$ & 787.0/6 & 0.00 & 0.00 \\
\hline
$\{c_3,c_4\}$ & $11.1100(20)$ & 2.71/5 & 0.75  & 0.26 \\
$\{c_3,c_5\}$ & $11.1094(12)$ & 2.7/5 & 0.75  & 0.26 \\
$\{c_3,c_6\}$ & $11.1102(23)$ & 2.71/5 & 0.74  & 0.26 \\
\hline
\end{tabular}
\end{minipage}%
\begin{minipage}{0.5\linewidth}%
\begin{tabular}{lcccc}
\hline
fit ansatz & $f_\mathrm{c}$ & $\chi^2/\mathrm{dof}$ & $p$ & $w_\mathrm{AIC}$ \\
\hline
base model & $10.99910(60)$ & 5760.0/6 & 0.00 & 0.00 \\
\hline
$\{c_3\}$ & $11.1099(16)$ & 2.67/5 & 0.75 & 0.43 \\
$\{c_4\}$ & $11.04207(85)$ & 514.0/5 & 0.00 & 0.00 \\
$\{c_5\}$ & $11.0747(12)$ & 35.4/5 & 0.00 & 0.00 \\
$\{c_6\}$ & $11.0676(11)$ & 132.0/5 & 0.00 & 0.00 \\
\hline
$\{c_3,c_4\}$ & $11.1095(31)$ & 2.64/4 & 0.62  & 0.16 \\
$\{c_3,c_5\}$ & $11.1114(64)$ & 2.61/4 & 0.63  & 0.16 \\
$\{c_3,c_6\}$ & $11.1095(39)$ & 2.65/4 & 0.62  & 0.16 \\
\hline
\end{tabular}
\end{minipage}%
\caption{\label{tab:lambda_max_0.7_0.5}
Same as Tab. \ref{tab:lambda_max_1.0}, but pruned and for $\hat{\lambda}_\mathrm{max}=0.7$ (left) and $\hat{\lambda}_\mathrm{max}=0.5$ (right). The pruning acts whenever $\#\mathrm{dof}$ is reduced; out of the ans\"atze with $\mathrm{dof}=6$ only the one with $p=0.83$ is augmented by another parameter. The AIC-weights will be normalized later; see text for details.}
\end{table}

\begin{table}[!tb]
\begin{minipage}{0.5\linewidth}%
\begin{tabular}{lcccc}
\hline
fit ansatz & $f_\mathrm{c}$ & $\chi^2/\mathrm{dof}$ & $p$ & $w_\mathrm{AIC}$ \\
\hline
base model & $11.03216(80)$ & 1690.0/5 & 0.00 & 0.00 \\
\hline
$\{c_3\}$ & $11.1095(21)$ & 2.58/4 & 0.63 & 0.27 \\
$\{c_4\}$ & $11.04444(87)$ & 252.0/4 & 0.00 & 0.00 \\
$\{c_5\}$ & $11.0742(13)$ & 34.6/4 & 0.00 & 0.00 \\
$\{c_6\}$ & $11.0738(13)$ & 56.1/4 & 0.00 & 0.00 \\
\hline
$\{c_3,c_4\}$ & $11.1106(42)$ & 2.50/3 & 0.48 & 0.11 \\
$\{c_3,c_5\}$ & $11.1110(65)$ & 2.52/3 & 0.47 & 0.10 \\
$\{c_3,c_6\}$ & $11.1109(52)$ & 2.50/3 & 0.48 & 0.11 \\
\hline
\end{tabular}
\end{minipage}%
\begin{minipage}{0.5\linewidth}%
\begin{tabular}{lcccc}
\hline
fit ansatz & $f_\mathrm{c}$ & $\chi^2/\mathrm{dof}$ & $p$ & $w_\mathrm{AIC}$ \\
\hline
base model & $11.05153(96)$ & 286.0/4 & 0.00 & 0.00\\
\hline
$\{c_3\}$ & $11.1108(36)$ & 2.4/3 & 0.49 & 0.18 \\
$\{c_4\}$ & $11.0334(26)$ & 233.0/3 & 0.00 & 0.00 \\
$\{c_5\}$ & $11.0824(21)$ & 9.18/3 & 0.03 & 0.01 \\
$\{c_6\}$ & $11.0873(24)$ & 7.88/3 & 0.05 & 0.01 \\
\hline
$\{c_3,c_4\}$ & $11.1079(55)$ & 1.95/2 & 0.38 & 0.08 \\
$\{c_3,c_5\}$ & $11.1074(96)$ & 2.26/2 & 0.32 & 0.07 \\
$\{c_3,c_6\}$ & $11.1068(84)$ & 2.14/2 & 0.34 & 0.08 \\
\hline
\end{tabular}
\end{minipage}%
\caption{\label{tab:lambda_max_0.35_0.25}
Same as Tab.~\ref{tab:lambda_max_1.0}, but pruned and for $\hat{\lambda}_\mathrm{max}=0.35$ (left) and $\hat{\lambda}_\mathrm{max}=0.25$ (right).}
\end{table}

Together, these three measures constitute a systematic procedure for generating and selecting the fitting ans\"atze that is appropriate to the data, but we emphasize that all numbers quoted so far refer to the fixed choice $\hat{\lambda}_\mathrm{max}=1.0$.

\begin{figure}[!tb]
\centering
\includegraphics[width=0.5\linewidth]{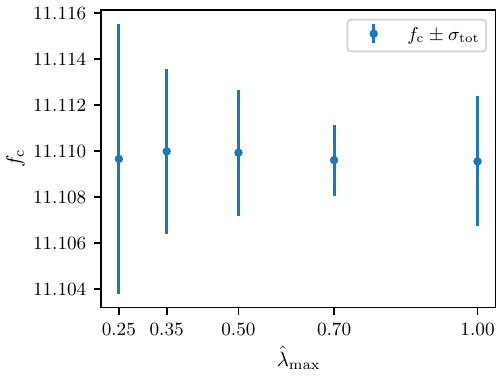}%
\includegraphics[width=0.5\linewidth]{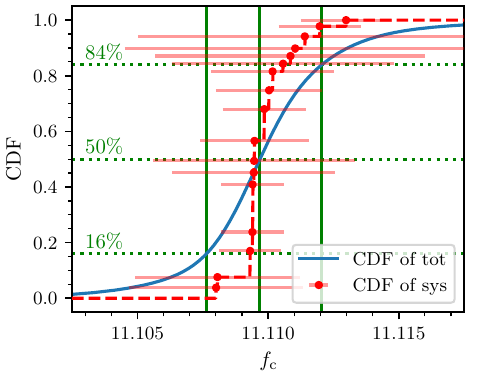}%
\caption{\label{fig:fc_final}
Left: Values of $f_\mathrm{c}$ obtained with the procedure shown in the right panel of Fig.~\ref{fig:fc_lambdamax_1}, separately for $\hat{\lambda}_\mathrm{max} \in \{1.0, 0.7, 0.5, 0.35, 0.25\}$. Right: CDF of all statistically acceptable fits from the left panel except for $\lambda_\mathrm{max}=0.25$, merging results from the other four choices of $\hat{\lambda}_\mathrm{max}$ in the same manner as for a single $\hat{\lambda}_\mathrm{max}$ in Fig.~\ref{fig:fc_lambdamax_1}. From this combined analysis, we obtain our final result $f_\mathrm{c}=11.1097(20)_\mathrm{stat}(09)_\mathrm{sys}=11.1097(22)_\mathrm{tot}$.}
\end{figure}

The next step is thus to explore how things change if we omit our largest $\hat{\lambda}$ from the fits.
The answer for $\hat{\lambda}_\mathrm{max}=0.7$ and $\hat{\lambda}_\mathrm{max}=0.5$ is listed in Tab.~\ref{tab:lambda_max_0.7_0.5}, and the choices $\hat{\lambda}_\mathrm{max}=0.35,0.25$ are documented in Tab.~\ref{tab:lambda_max_0.35_0.25}.
The good news is that we get good $p$-values (for some of the fits) with all these choices.
A survey of the central values and total uncertainties obtained with these five choices of $\hat{\lambda}_\mathrm{max}$ is shown in the left panel of Fig.~\ref{fig:fc_final}.
The central value seems remarkably stable under a variation of $\hat{\lambda}_\mathrm{max}$, and the total uncertainty is minimal at $\hat{\lambda}_\mathrm{max}=0.7$.
If we were to go for an aggressive uncertainty quantification, we would select the choice $\hat{\lambda}_\mathrm{max}=0.7$.
To come up with a conservative estimate, we feel it is appropriate to include the four choices $\hat{\lambda}_\mathrm{max}=0.35,0.5,0.7,1.0$ in the final analysis.
The CDF of all acceptable fits with these $\hat{\lambda}_\mathrm{max}$ is shown in the right panel of Fig.~\ref{fig:fc_final}.
In the way described above, we ``smooth out'' the $17$-step function by replacing each step with a normal cumulative distribution (depicted by the full blue line).
From the positions of the $16^\mathrm{th}$ and $84^\mathrm{th}$ percentiles we obtain our final result $f_\mathrm{c}=11.1097(20)_\mathrm{stat}(09)_\mathrm{sys}=11.1097(22)_\mathrm{tot}$.


\section{Comparisons and Conclusions\label{sec:conc}}

\begin{table}[!tb]
\centering
\begin{tabular}{c|c|c}
\hline
Method & $f_\mathrm{c}$ & Year \\
\hline
Monte Carlo~\cite{Loinaz:1997az}  & 10.24(3) & 1998\\
Monte Carlo~\cite{Schaich:2009jk} & $10.8^{+0.10}_{-0.05}$ & 2009\\
Monte Carlo (SLAC derivative)~\cite{Wozar:2011gu} & 10.92(12) & 2012 \\
Matrix product states~\cite{Milsted:2013rxa}& 11.077(8), 11.050(3) & 2013\\
Monte Carlo~\cite{Bosetti:2015lsa} & 11.15(7) & 2015\\
Borel RPT~\cite{Serone:2018gjo} & 11.22(13) & 2018\\
Monte Carlo~\cite{Bronzin:2018tqz} & 11.055(24) & 2019\\
Tensor network~\cite{Kadoh:2018tis} & 10.913(56) & 2019\\
Tensor network~\cite{Delcamp:2020hzo} & 11.0861(90) & 2020\\
Optimized perturbation theory~\cite{Heymans:2021rqo} & 11.116(10) & 2021\\
Tensor network~\cite{Vanhecke:2021noi} & 11.09698(31) & 2022\\
This work & 11.1097(22) & 2025\\
\hline
\end{tabular}
\caption{\label{tab:literature}
Comparison with the literature, always quoting the total uncertainty.
Out of these references Loinaz and Willey \cite{Loinaz:1997az}, Schaich and Loinaz \cite{Schaich:2009jk}, Bosetti et al.\ \cite{Bosetti:2015lsa} and Bronzin et al.\ \cite{Bronzin:2018tqz} use the same methodology as we use in this work.}
\end{table}

\begin{figure}
    \centering
    \includegraphics[width=\linewidth]{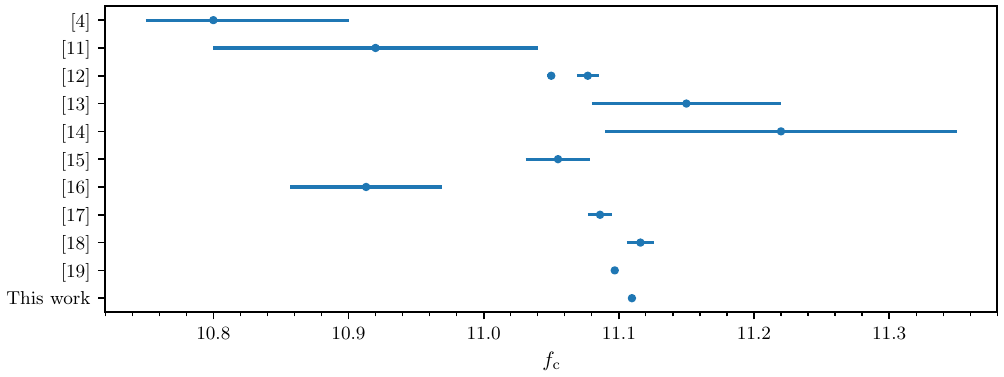}
    \caption{Comparison with literature values (see Tab.~\ref{tab:literature}); in some cases the error bar is smaller than the size of the symbol.
    The result of Loinaz and Willey \cite{Loinaz:1997az} is not show, as it lies far outside the plotted range.}
    \label{fig:literature}
\end{figure}

In Tab.~\ref{tab:literature} and Fig.~\ref{fig:literature} we compare our result to earlier results in the literature.
Ref.~\cite{Loinaz:1997az} was the pioneering Monte Carlo study with lattice regularization, and they used a linear ansatz to extrapolate $f_\mathrm{c}(\hat{\lambda})$ to $\hat{\lambda}=0$.
Ref.~\cite{Schaich:2009jk} was the first paper that reached the precision necessary to demonstrate that their data are incompatible with a purely linear behavior, and a term $\hat{\lambda}\ln(\hat{\lambda})$ is needed.
As a byproduct, this logarithmic contribution would significantly increase the best estimate of $f_\mathrm{c}=\lim_{\hat{\lambda}\to0}f_\mathrm{c}(\hat{\lambda})$.
From a conceptual point of view Ref.~\cite{Wozar:2011gu} is interesting, since they use the SLAC derivative to discretize the action (\ref{def_orig}).
This leads to a non-local theory, and to different results at finite $\hat{\lambda}$.
However, in the limit $\hat{\lambda}\to0$ they seem to reach the same \emph{universal} value $f_\mathrm{c}$ as with the standard discretization (but we are unaware of a formal proof).
Refs.~\cite{Bosetti:2015lsa,Bronzin:2018tqz} provide a continuation of the previous methodology.

More recently, dedicated perturbative techniques \cite{Serone:2018gjo,Heymans:2021rqo} and papers based on the tensor network approach \cite{Kadoh:2018tis,Delcamp:2020hzo,Vanhecke:2021noi} seem to reach very high (claimed) precision.
An inconvenient observation is that our value is inconsistent with Ref.~\cite{Vanhecke:2021noi}, at the level of $5.7\sigma$ (combined uncertainty).
After investigating the issue with diligence, we suspect that the root cause is not the framework for generating the data (Monte Carlo versus Tensor Network) but rather the family of admitted fitting functions.
Unfortunately, we were unable to repeat their analysis, since we find only scatter plots and no aggregate data in their paper.
But it strikes us that they do not consider the term $\hat{\lambda}\ln^2(\hat{\lambda})$ in their function base.
This is precisely the term with the coefficient $c_3$ in Tab.~\ref{tab:lambda_max_1.0} that was decisive for getting a good $p$-value (and similarly in Tabs.~\ref{tab:lambda_max_0.7_0.5}, \ref{tab:lambda_max_0.35_0.25}).
However, we can test how our result would change if we were to exclude this term from our fits.
In Tab.~\ref{tab:lambda_max_1.0} the fit with $c_4,c_5,c_6$ would give $f_\mathrm{c}=11.0938(24)$ with $p=0.23$, while all fits with only two of these three terms do not work.
In subsequent tables this fit does not show up again, due to the pruning procedure, but it strikes us that our data produce a number similar to theirs if we exclude the $\hat{\lambda}\ln^2(\hat{\lambda})$ term from the analysis.
As stated in Sec.~\ref{sec:fc} we are unaware of an argument why this term should be excluded, and we observe that including the $\hat{\lambda}\ln^2(\hat{\lambda})$ term has a tremendously beneficial effect on the fit quality.
We suspect that it will be enlightening if the authors of Ref.~\cite{Vanhecke:2021noi} redo their analysis with this function included, and we think that there is good hope that this will eliminate (or at least diminish) the disagreement.

For the time being, we stress that we processed our data with an analysis strategy that is theoretically well founded (see \cite{Akaike:1974vps,BMW:2014pzb,Borsanyi:2020mff} and references therein) and pragmatic in the sense that it penalizes fitting ans\"atze for not describing the data, but not for an a-priori reason.
The lattice field theory approach introduces two systematic effects (finite volume and finite lattice spacing) that we carefully extrapolated away, with full control over the statistical and systematic uncertainties involved.


\subsection*{Acknowledgments}

Computations per performed on (the remnants of) a small PC cluster at the University of Wuppertal that was originally financed by DFG as part of the SFB-TRR-55 grant.


\appendix

\section{Evaluating the regularized tadpole integral in terms of Bessel functions\label{sec:app}}

The tadpole integral, regularized with lattice spacing $a$, takes the form
\begin{equation}
    A(\hat{\mu}^2;a)=\frac{1}{N^d}\sum_{k\in\tilde{\Lambda}}\frac{1}{\hat{k}^2+\hat{\mu}^2}
    \stackrel{N\to\infty}{=}\int_{-\pi}^{\pi}\frac{\mathrm{d}^dk}{(2\pi)^d}\frac{1}{\hat{k}^2+\hat{\mu}^2}\;.
\end{equation}
We insert the Schwinger parameterization
\begin{equation}
    \frac{1}{A}=\int_0^\infty\mathrm{d}t~e^{-tA}
\end{equation}
for the integrand and find
\begin{equation}
    A(\hat{\mu}^2;0)=\int_0^\infty\mathrm{d}t~e^{-t\hat{\mu}^2}\int_{-\pi}^{\pi}\frac{\mathrm{d}^dk}{(2\pi)^d}~e^{-t\hat{k}^2}
    \label{app_one}
\end{equation}
where the $d$ dimensional integration over $k$ can be factorized as
\begin{equation}
    \int_{-\pi}^{\pi}\frac{\mathrm{d}^dk}{(2\pi)^d}~e^{-t\hat{k}^2}=\left[\int_{-\pi}^\pi\frac{\mathrm{d}k}{2\pi}~e^{-4t\sin^2(k/2)}\right]^d\;.
\end{equation}
Using $\sin^2(k/2)=(1-\cos(k))/2$ this is rewritten as
\begin{equation}
    \int_{-\pi}^{\pi}\frac{\mathrm{d}k}{2\pi}~e^{-4t\sin^2(k/2)}=\int_{-\pi}^{\pi}\frac{\mathrm{d}k}{2\pi}~e^{-2t}e^{2t\cos(k)}=e^{-2t}I_0(2t)
    \label{app_two}
\end{equation}
where $I_0$ is the modified Bessel function of the first kind in the family
\begin{equation}
    I_n(t)=\frac{1}{\pi}\int_0^\pi\mathrm{d}\theta~e^{t\cos(\theta)}\cos(n\theta)\;.
\end{equation}
Combining (\ref{app_one}) and (\ref{app_two}) gives the final form
\begin{equation}
    A(\hat{\mu}^2)=\int_0^\infty\mathrm{d}t~e^{-t\hat{\mu}^2}\left[e^{-2t}I_0(2t)\right]^d
\end{equation}
that was used in the implicit definition (\ref{implicit_1}, \ref{implicit_2}).


\section{Assessment of statistical and systematic uncertainties\label{sec:app_b}}

This procedure follows the analysis introduced in Ref.~\cite{BMW:2014pzb} with the refinements of Ref.~\cite{Borsanyi:2020mff}.
We summarize it here for completeness and to provide an easily accessible reference for the method used.
To estimate the statistical and systematic components of the total uncertainty, we treat the set of AIC-weighted fits as a Gaussian mixture.
Each fit $i$ contributes a Gaussian distribution with central value $f_{\mathrm{c},i}$ and statistical width $\sigma_i$, weighted by $w_i$.
The resulting probability density is
\begin{equation}
    \sum_i w_i\, N(y; f_{\mathrm{c},i}, \sigma_i),
\end{equation}
which corresponds to the cumulative distribution function
\begin{equation}
    P(y;\kappa)=\int_{-\infty}^{y}\mathrm{d}y^\prime~\sum_i w_i\,N(y^\prime; f_{\mathrm{c},i}, \sigma_i\sqrt{\kappa})\;,
\end{equation}
where the parameter $\kappa$ rescales the statistical widths without modifying the spread of the central values.
The total uncertainty is defined from the $16^{\mathrm{th}}$ and $84^{\mathrm{th}}$ percentiles of the CDF evaluated at $\kappa=1$,
\begin{equation}
    \sigma^2_{\mathrm{tot}} \equiv \left[\frac{1}{2}(y_{84}-y_{16})\right]^2, \qquad\text{with}\qquad P(y_{16};1)=0.16,\quad P(y_{84};1)=0.84\;.
\end{equation}
This uncertainty is decomposed into statistical and systematic contributions with
\begin{equation}
    \sigma_{\mathrm{stat}}^2 + \sigma_{\mathrm{sys}}^2 \equiv \sigma_{\mathrm{tot}}^2\;.
\end{equation}
To isolate the statistical component, we artificially rescale the statistical widths by a
factor $\sqrt{\kappa}$ and extract the corresponding percentile interval
\begin{equation}
    \kappa\sigma_{\mathrm{stat}}^2 + \sigma_{\mathrm{sys}}^2 \equiv \left[\frac{1}{2}(\tilde{y}_{84}-\tilde{y}_{16})\right]^2 \qquad\text{with}\qquad P(\tilde{y}_{16};\kappa)=0.16,\quad P(\tilde{y}_{84};\kappa)=0.84\;.
\end{equation}
Considering the difference between this expression with arbitrary $\kappa$ and evaluated at $\kappa=1$ yields
\begin{equation}
    \left[\frac{1}{2}(\tilde{y}_{84}-\tilde{y}_{16})\right]^2-\left[\frac{1}{2}(y_{84}-y_{16})\right]^2 = (\kappa-1)\sigma_\mathrm{stat}^2
\end{equation}
which provides a direct relation between the change in percentile width under the rescaling and the underlying statistical variance.
For our analysis we choose $\kappa=2$.


\end{document}